\documentclass[preprint,showpacs,preprintnumbers,amsmath,amssymb]{revtex4}%
\usepackage{graphicx}
\usepackage{dcolumn}
\usepackage{bm}
\usepackage{amsmath}
\usepackage{amsfonts}
\usepackage{amssymb}%
\providecommand{\U}[1]{\protect\rule{.1in}{.1in}}
\newcommand{\be}{\begin{equation}}
\newcommand{\en}{\end{equation}}
\newcommand{\bea}{\begin{eqnarray}}
\newcommand{\ena}{\end{eqnarray}}
\begin{document}
\title{Warm intermediate inflation in the Randall-Sundrum II model
in the light of Planck 2015 and BICEP2 results:
 A
general dissipative coefficient }
\author{Ram\'on Herrera}
\email{ramon.herrera@ucv.cl}
\affiliation{Instituto de F\'{\i}sica, Pontificia Universidad Cat\'{o}lica de
Valpara\'{\i}so, Avenida Brasil 2950, Casilla 4059, Valpara\'{\i}so, Chile.}
\author{Nelson Videla}
\email{nelson.videla@ing.uchile.cl}
\affiliation{Departamento de F\'{\i}sica, FCFM, Universidad de Chile, Blanco Encalada 2008, Santiago, Chile}
\author{Marco Olivares}
\email{marco.olivaresr@mail.udp.cl}
\affiliation{Facultad de Ingenier\'ia, Universidad Diego Portales,
Avenida Ej\'ercito Libertador 441, Casilla 298-V, Santiago, Chile.}
\date{\today}

\begin{abstract}
A warm inflationary Universe in the Randall-Sundrum II model
during intermediate inflation is studied. For this purpose, we
consider a general form for the dissipative coefficient
$\Gamma(T,\phi)=C_{\phi}\,\frac{T^{m}}{\phi^{m-1}}$, and also
analyze  this inflationary model in the weak and strong
dissipative regimes. We study the evolution of the Universe under
the slow-roll approximation and find solutions to the full
effective Friedmann equation in the brane-world framework. In
order to constrain the parameters in our model, we consider the
recent data from the BICEP2-Planck 2015 data together with the
necessary
 condition for warm inflation $T>H$, and also
 the condition from the weak (or strong) dissipative regime.

\end{abstract}

\maketitle

\section{Introduction}

In modern cosmology, our notions concerning the physics of the
very early Universe have introduced a new element, the
inflationary period \cite{R1,R102,R103,R104,R105,R106}, which
provides an attractive approach for solving some problems of the
standard Big-Bang model, like the flatness, horizon, monopoles,
among other. However, the essential feature of inflation is that
can generate a novel mechanism to explain the Large-Scale
Structure (LSS) of the Universe \cite{R2,R202,R203,R204,R205} and
provide a causal interpretation of the origin of the anisotropies
observed in the Cosmic Microwave Background (CMB)
radiation\cite{astro,astro2,astro202,Hinshaw:2012aka,Ade:2013zuv,Ade:2013uln,Planck2015}.

By other hand, with respect to the dynamical mechanisms of
inflation, the warm inflation scenario, as opposed to the standard
cold inflation, has the attractive feature that it avoids the
reheating period at the end of the accelerated expansion, because
of the decay of the inflaton field into radiation and particles
during the slow-roll phase \cite{warm}. During the warm inflation
scenario the dissipative effects are important, since radiation
production occurs concurrently together with the inflationary
expansion. The dissipating effects arise from a friction term
which describes the process of the scalar field decaying into a
thermal bath.  Also, during the scenario of  warm inflation the
thermal fluctuations arises from the inflationary epoch may play a
fundamental  function in producing the initial fluctuations
indispensable  for the Large-Scale Structure (LSS) formation
\cite{62526,1126}. An indispensable condition for warm inflation
scenario to occur is the presence of a radiation field with
temperature $T>H$ during the inflationary expansion of the
Universe. Since the thermal and quantum fluctuations are
proportional to $T$ and $H$, respectively \cite{warm,62526,1126},
so when $T>H$,  the thermal fluctuations of the inflaton field
predominates over the quantum fluctuations. Also,
inflation ends when the Universe heats up to   become radiation
dominated, and then the Universe smoothly enters to the
radiation dominated era, without the need of a reheating scenario
\cite{warm}. For a  review of warm inflation, see e.g. Ref.
\cite{Berera:2008ar}.

On the other hand, cosmological implications of string/M-theory
have currently attracted a great deal of attention; specifically
some were concerned with brane-antibrane configurations such as
space-like branes \cite{sen1}. In this configuration, the standard
model of particles is confined to the brane, while gravitation
propagates into the bulk space-time. Here, the effect of extra
dimensions induces extra terms in the Friedmann equation
\cite{1,3,8}. In particular, the cosmological Randall-Sundrum II
model (RS II) has received great attention in the last
years\cite{RS}. Randall and Sundrum suggested  two similar but phenomenologically different
 brane world scenarios. In this form,
there are two versions of the Randall-Sundrum model,
 generally mentioned  as Randall-Sundrum I (RS1)\cite{RS1} and
 Randall-Sundrum II (RS2)\cite{RS}. Somewhat confusingly, the RS1 model
 includes two branes, while  the RS2 model only contains a single brane.

In the RS1 model, we have two 3-branes separated by a region of
five-dimensional anti-de Sitter spacetime. Here, the fifth
coordinate is compactified on $S^1/Z_2$, and the branes have
equal;  positive and negative tension branes are on the two
specified points. In this model of brane the matter fields are
confined on the two brane and the gravity propagates  in the
5-dimensional bulk. However, one important point of discussion in
this brane model, is the mechanism that works to select the
necessary separation distance between the two branes  called
radius stabilization\cite{est} (see also Ref.\cite{est2}).
Similarly, the stabilization mechanism plays a crucial role in the
recuperation of 4-dimensional Einstein gravity\cite{rec}.

In contrast, the RS2 model\cite{RS} contains a single, positive
tension brane and a non-compact extra dimension is infinite in
extent and the radius stabilization
 is not present in this brane model. By other hand, in the RS2 model, the observable
 universe is a four-dimensional time-like hypersurface. The five-dimensional  energy-momentum
 tensor can be splited in a regular and a distributional part. The regular part describes the non-standard
 matter fields in five dimension and the distributional term containing the brane tension and the standard
 matter fields on the brane. Except gravitation, all standard model interactions and matter fields are confined to the
 brane.

These alternatives to Einstein's General Relativity cosmology
belongs to the so called brane-world cosmological models. For a
comprehensible  review of brane-world cosmology, see e.g.
Refs.\cite{4,5,M}. In the observational aspect, nowadays there is
strong evidence that the very early Universe could have
experimented an inflationary expansion period, as was pointed out
in the begging of the introduction. An important feature of the
inflationary period, is that is located in a period of
cosmological evolution where the effects predicted by
string/M-theory are relevant. For this reason, over the last
decade, there has been great interest in the construction of
inflationary models inspired in these theories. In the
following we will concentrate only in the RS2 model, which forms
the basis for the rest of this work.

As regards exact solutions, one the most interesting in the
inflationary Universe can be found  by using an exponential
potential, which is often called a power-law inflation, since the
scale factor has a power-law-type evolution, here the scale factor
is given by  $a(t)\sim t^{p}$, in which $p>1$\cite{power}. Also,
an exact solution can be obtained in the de Sitter inflationary
Universe, where a constant effective  potential is considered; see
Ref.\cite{R1}. Moreover, exact solutions can also be found  for
the scenario of intermediate inflation, for which the scale factor
$a(t)$ evolves as
\begin{equation}
a(t)=\exp[\,A\,t^{f}],  \label{at}
\end{equation}
where $A$ and $f$ are two constants; $A>0$ and $0<f<1$
\cite{Barrow1}. It is well know that the expansion rate in
this model is slower than de Sitter inflation, but faster than
power-law inflation; this is why it is known as "intermediate".
This inflationary model  was originally developed as an exact
solution, however, intermediate inflation may be described
from the slow-roll analysis. Under the slow-roll approximation the
scalar spectral index $n_s\sim 1$, and  for the specific value of
$f$ given by $f=2/3$ the spectral index is $n_s=1$
(Harrizon-Zel'dovich spectrum)
 \cite{Barrow2}. Also, an
important observational quantity in this model, is that the
tensor-to-scalar ratio $r$, which is significantly $r\neq
0$\cite{ratior,Barrow3}. Other motivation to consider this type of
expansion for the scale factor comes from string/M-theory, which
appears to be relevant in the low-energy string effective
action{\cite{KM,ART}} (see also,
Refs.\cite{BD,Varios1,Varios2,Sanyal}). These theories can be
utilized to resolve the initial singularity and describe the
present acceleration in the universe, among others\cite{new}.
Also, the approach of consider the warm intermediate inflation has
already been studied in the literature,\cite{Herrera:2014nta} in
the context of other frameworks.


In this way, the goal of this paper is to analyze the possibility
that a higher dimensional scenario, in particular the RS II
brane-world model, can describe the dynamics of the Universe in
its very early epochs. We propose this possibility in the context
of warm inflation scenario for a Universe evolving according to
the intermediate scale factor, and how  a generalized form of
dissipative coefficient $\Gamma(T,\phi)\propto T^{m}/\phi^{m-1}$
influences the dynamics of our model.  In order to study our
brane-warm intermediate model we will consider the full effective
Friedmann equation, and not the lower energy limit or the high
energy limit for the effective Friedmann equation. Also, we will
 study the cosmological perturbations, which are
 written in terms of several parameters. Here, the parameters are constrained from
 the
 BICEP2 data\cite{BICEP} together with the Planck satellite\cite{Ade:2013uln} and Planck 2015\cite{Planck2015}.
 By other hand, it is well known that
 the BICEP2 experiment data has important
 theoretical significance on the  amplitude of
 primordial gravitational waves produced during inflation. In this form,
  the tensor-to-scalar ratio $r$ from the BICEP2 data, has
 been found at more than 5$\sigma$ confidence level (C.L.)  in which the ratio
 $r=0.2_{-0.05}^{+0.07}$ at 68$\%$ C.L., and with foreground subtracted
 $r=0.16_{-0.05}^{+0.06}$\cite{BICEP}. Nevertheless, this value of the tensor-to-scalar ratio has
 become less transparent , see e.g.  Ref.\cite{Prob1}. Of this way, a detailed analysis from Planck
  and BICEP2 data would be necessary for a definitive answer
 to the diffuse Galactic dust contamination. Recently, the Planck collaboration published new data of
 enhanced precision of the CMB anisotropies\cite{Planck2015}.
  Here, the Planck full mission data improved the upper bound on the tensor to scalar ratio $r_{0.002}<0.11 (95\% C.L.)$, and this upper bound for the ratio $r$
  is similar to obtained
  from Refs.\cite{BICEP,Ade:2013uln}, in which $r<0.12 (95\% C.L.)$.

The outline of the paper is as follows: The next section presents a short
review of the effective Friedmann equation for the Randall-Sundrum type II scenario. In
the section III we study the dynamics of warm inflation in this brane-world model,
in the weak and strong dissipative regimes; specifically,  we
obtain explicit expressions for the inflaton scalar field, dissipative coefficient, and effective scalar
potential. Immediately, we compute the cosmological perturbations in both dissipative regimes,
obtaining expressions for observational quantities such as the scalar power spectrum, scalar spectral index, and the tensor-to-scalar ratio.
 At the end, section IV summarizes our results and exhibits our conclusions. We chose units so that
$c=\hbar=1$.

\section{The Brane-Intermediate warm inflation scenario}

Followings, Refs.\cite{3,8} we consider a five-dimensional
scenario, for which the modified Friedmann equation for a spatially
flat Friedmann-Robertson-Walker (FRW) metric,  has the form

\begin{equation}
H^2=\kappa\,\rho\left[1+\frac{\rho}{2\tau}\right]+\frac{\Lambda_4}{3}+\frac{\xi}{a^4},
\label{eq1}
\end{equation}
where  $H=\dot{a}/a$ represents  the Hubble
parameter, $a$ denotes the scale factor, the constant $\kappa=8\pi
G/3=8\pi/3m_p^2$, and  the
energy density $\rho$ corresponds to the energy density of the matter content confined to the brane, $\Lambda_{4}$
is the effective four-dimensional cosmological constant on the brane, and $\xi/a^4$ denotes the influence
of the bulk gravitons on the brane, where $\xi$ is an integration
constant. The term $\tau$ represents the
brane tension and, considering  the nucleosynthesis epoch, the value of the brane
tension  is constrained to be $\tau
>$ (1MeV)$^4$ \cite{Cline}.  However,  a stronger limit for the value of the brane
tension results  from usual tests for deviation from Newton`s
law, for which $\tau\geq $(10
 TeV)$^4$, see Ref.\cite{Bla}.

In the following, we will consider that the cosmological constant $\Lambda_{4}=0$, and
once inflation begins, the term
$\xi/a^4$ rapidly becomes unimportant. In this form,  the effective Friedmann equation
 given by Eq.(\ref{eq1}) becomes
\begin{equation}
H^{2}=\kappa\,\rho\,\left[1+{\rho \over 2\,\tau}\right].
\label{HC}%
\end{equation}

On the other hand, during   warm inflation the Universe is filled with a
self-interacting scalar field  with  energy density $\rho_{\phi}$ together with
 a radiation field with energy density  $\rho_{\gamma}$.
In this form, the total energy density $\rho$ of the Universe can be written as
 $\rho=\rho_{\phi}+\rho_{\gamma}$. Here,
 the energy density $\rho_{\phi}$ and the pressure $P_{\phi}$ of the scalar field are
given by
$\rho_{\phi}=\dot{\phi}^{2}/2+V(\phi)$ and $P_{\phi}=\dot{\phi}^{2}/2-V(\phi)$, respectively.
 The term $V(\phi)$ represents
the effective scalar potential. In the following, we will consider that the dots
 mean derivatives with respect to cosmic time.

We will assume that the total energy density $\rho$ is confined in the brane,
and then the continuity equation for the total energy density becomes
$\dot{\rho}+3\,H\,(\rho+P)=0$.  In this way, following Ref.\cite{warm}, the dynamical equations for
$\rho_{\phi}$ and $\rho_{\gamma}$ during  warm inflation
can be written as
\begin{equation}
\dot{\rho_{\phi}}+3\,H\,(\rho_{\phi}+P_{\phi})=-\Gamma\;\;\dot{\phi}^{2},
\label{key_01}%
\end{equation}
and
\begin{equation}
\dot{\rho}_{\gamma}+4H\rho_{\gamma}=\Gamma\dot{\phi}^{2}. \label{key_02}%
\end{equation}
Here, $\Gamma>0$ represents
 the dissipative coefficient and describes the process of scalar field decaying into radiation
during the inflationary expansion \cite{warm}. In the context of
brane-warm inflationary model, in Ref.\cite{cid} was studied a
high energy scenario during the strong dissipative
 regime.

 From quantum field theory, the dissipative coefficient $\Gamma$
was computed in a supersymmetric model for a low-temperature
scenario \cite{26}. In particular,  for a scalar field
with multiplets of heavy and light fields, it is possible to obtain several
expressions for the  dissipative coefficient $\Gamma$, see e.g.,
\cite{26,28,2802,Zhang:2009ge,BasteroGil:2011xd,BasteroGil:2012cm}.

Following Refs.\cite{Zhang:2009ge,BasteroGil:2011xd}, we consider
a general form for the dissipative coefficient
$\Gamma(T,\phi)$ given by

\begin{equation}
\Gamma(T,\phi)=C_{\phi}\,\frac{T^{m}}{\phi^{m-1}}, \label{G}%
\end{equation}
where  the constant  $C_\phi$ is related   with  the dissipative
microscopic dynamics and the exponent $m$ is an integer. This expression for
 the dissipative coefficient includes different cases, depending of the
 values of $m$, see Refs.
\cite{Zhang:2009ge,BasteroGil:2011xd}. Concretely , for the
special value of $m=3$, for which $\Gamma= C_\phi T^3\phi^{-2}$, the parameter
$C_\phi$ agrees with  $C_{\phi}=0.02\,h^{2}\,\mathcal{N}_Y$,
where a generic supersymmetric model with chirial superfields
$\Phi$, $X$ and $Y_i$, $i=1,...\mathcal{N}_Y$ is studied
\cite{BasteroGil:2012cm,new2}. For the special case $m=1$,
the dissipative coefficient  is related with the high temperature supersymmetry
(SUSY) case.  Finally, for the cases $m=0$ and $m=-1$, the term $ \Gamma(T,\phi)$
 represents an exponentially decaying propagator in
the high temperature SUSY model and
 the non-SUSY
case, respectively\cite{28,PRD}.

Considering that in the scenario of
warm inflation the energy density $\rho_{\phi}$ predominates  over the density $\rho_{\gamma}$
\cite{warm,62526},
 then the Eq.(\ref{HC}) becomes
\begin{equation}
H^{2}\approx \kappa\,\rho_{\phi}\,\left(1+{\rho_{\phi} \over 2\,\tau}\right)
=\kappa\,\left(\frac{\dot{\phi}^2}{2}+V(\phi)\right)\left[1+{\dot{\phi}^2+2V(\phi)\over
4\,\tau}\right].\label{HH}
\end{equation}

In the following, we will not study the effective Friedmann
equation, given by Eq.(\ref{HH}), in the lower energy limit i.e.,
$\rho_\phi\ll\tau$ or in the high energy limit i.e.,
$\rho_\phi\gg\tau$  as our starting point, instead we will
consider the full effective Friedmann Eq.(\ref{HH}). Here, we
note that there are two ways of deriving the Friedmann's  equation
from five-dimensional  Einstein's equations. The first method is
rather simple and considers only the bulk equations. The second
approach   utilizes the geometrical relationship between
four-dimensional and five-dimensional quantities. However, the
Einstein   equations and in particular the Friedmann's equation in
the bulk include different functions (from FRW metric in
5-dimensional). In particular, these functions are subjects to
conditions (junction conditions) on the brane localized at $y=0$
and symmetry ($Z_2$-symmetry) when integrate over $y$. In this
form, we could not obtain analytical solutions considering  the
full 5-dimensional equations of motion from 4-dimensional
analytical solutions. In the following,  we will obtain analytical
solutions in four-dimensional of the full effective Friedmann's
equation only.

In this form, solving the cuadratic equation (\ref{HH}) for
$\rho_{\phi}$ (where we take the solution for which
$\dot{\rho}_{\phi}<0$), and combining with Eqs. (\ref{key_01}),
results
\begin{equation}
\dot{\phi}^{2}= {\frac{2 }{3\kappa}}\frac{(-\dot{H})}{(1+R)}
\,\left[1+{2\,H^2 \over \kappa\,\tau}\right]^{-1/2}, \label{inf3}%
\end{equation}
where  the quantity $R$ denotes the ratio between $\Gamma$ and the Hubble
parameter, defined as
\begin{equation}
R=\frac{\Gamma}{3H}. \label{rG}%
\end{equation}
In the following, we will consider that for the case of the weak or strong dissipation
regime, the ratio $R$ satisfies $R<1$ or $R>1$, respectively.

Also, following Refs.\cite{warm,62526}, we consider that during warm inflation
 the radiation
production is quasi-stable, i.e.,  $\dot{\rho
}_{\gamma}\ll4H\rho_{\gamma}$ and $\dot{\rho}_{\gamma}\ll\Gamma\dot{\phi}^{2}%
$. In this
form, combining Eqs.(\ref{key_02}) and (\ref{inf3}), the energy density of the radiation field results
\begin{equation}
\rho_{\gamma}=\frac{\Gamma\dot{\phi}^{2}}{4H}=\frac{\Gamma(-\dot{H})}{6\kappa H(1+R)}
\,\left[1+{2\,H^2 \over \kappa\,\tau}\right]^{-1/2}. \label{rh}%
\end{equation}

On the other hand, we assume that the energy density of the radiation field
is given by $\rho_{\gamma}=C_{\gamma}\,T^{4}$, where  the constant $C_{\gamma}%
=\pi^{2}\,g_{\ast}/30$ and the term $g_{\ast}$ represents the number of
relativistic degrees of freedom. Combining  the above expression for
the energy density $\rho_{\gamma}$ and Eq.(\ref{rh}),
we find that the temperature of the thermal bath $T$ yields
\begin{equation}
T=\left[  \frac{\Gamma\,(-\dot{H})}{6\,\kappa\,\,C_{\gamma}H\,(1+R)}\right]^{1/4}
\,\left[1+{2\,H^2 \over \kappa\,\tau}\right]^{-1/8}. \label{rh-1}%
\end{equation}

From Eqs.(\ref{HC}) and (\ref{inf3}),  the effective scalar
potential results

\begin{equation}
V=\tau\,\left(-1+\sqrt{1+{2\,H^2 \over \kappa\,\tau}}\right)
+\frac{\dot{H}}{3\kappa(1+R)}
\,\left(1+{2\,H^2 \over \kappa\,\tau}\right)^{-1/2}  .\label{pot}%
\end{equation}
Here, we note that the effective potential could be determined  explicitly in terms of the scalar
field $\phi$, in the weak (or strong) dissipative regime.

Now, combining  Eqs.(\ref{rh-1}) and (\ref{G})  we get
\begin{equation}
\Gamma^{{\frac{4-m }{4}}}=\,C_{\phi}\phi^{1-m} \left[
\frac{-\dot{H}}{6\kappa\, C_{\gamma}H(1+R)}\right]  ^{m/4}\,
\left(1+{2\,H^2 \over \kappa\,\tau}\right)^{-m/8}. \label{G1}%
\end{equation}
 We note that the above expression determines the dissipation coefficient in the weak
(or strong) dissipative regime in terms of the scalar field
$\phi$.

In the following, we will study our brane-model for the general form of
the dissipative coefficient, given by Eq.(\ref{G}), during intermediate inflation,
where the scale factor evolves according to Eq.(\ref{at}). We
will restrict ourselves to the cases $R<1$ (weak regime) and
$R>1$ (strong regime). Also, in the following, we will study the cases $m=3$, $m=1$,
$m=0$ and $m=-1$ corresponding to the dissipative coefficient $\Gamma=\Gamma(T,\phi)$.

\subsection{ The weak dissipative regime.\label{subsection1}}

Firstly, we consider that our brane-model evolves
agreeing  to the weak dissipative regime, i.e., $R<1$ (or
analogously $\Gamma<3H$). In this from, combining Eqs.(\ref{at})
and (\ref{inf3}), the solution for the standard scalar field $\phi
$ results
\begin{equation}
\phi (t)-\phi _{0}=\frac{B[t]}{K},  \label{fisol}
\end{equation}%
here the quantity $\phi(t=0)=\phi_0$ is an integration constant,
than can be assumed as  $\phi(t=0)=\phi_0=0$ (without loss of
generality), and the constant $K$ is specified by $
K= \sqrt{\frac{6\,\kappa\,(1-f)}{A\,f}}\left({ \kappa\,\tau\over 2A^2f^2}\right)^{\frac{1}{2(1-f)}}. %
$ The function $\mathcal{B}[t]$, represents the incomplete Beta
function \cite{Libro}, given by
$$
B[t]= B\left[-\frac{\kappa \tau\,t^{2(1-f)} }{2(Af)^{2}};\frac{1}{%
4(1-f)},\frac{3}{4}\right].
$$

From Eqs.(\ref{at}) and (\ref{fisol}), we find that the Hubble parameter $H$  as function of the
scalar field, results
$ H(\phi)=A\,f\,\left(B^{-1}[K\,\phi]\right)^{-(1-f)}$.  Here, the function $B^{-1}[K\,\phi]$
corresponds to the inverse of the incomplete Beta function $B[t]$\cite{Libro}.

The effective potential as
function of the  scalar field $\phi$, from
 Eq.(\ref{pot}) and considering that $V>\dot{\phi}^2/2$, results
\begin{equation}
V(\phi)\simeq\tau\,\left(-1+
\sqrt{1+{2\,A^2f^2\left(B^{-1}[K\,\phi]\right)^{-2(1-f)} \over \kappa\,\tau}}\right) .\label{pot11}%
\end{equation}

Considering that $R<1$, from Eq.(\ref{G1}), the  dissipative coefficient $\Gamma$ in terms of the scalar
field becomes
\begin{equation}
\Gamma(\phi)=\,C_{\phi}^{{\frac{4}{4-m}}}\left[  \frac{1-f}%
{6\kappa\,C_{\gamma}\,B^{-1}[K\,\phi]}\right]^{\frac{m}{4-m}}\phi^{{\frac{4(1-m) }{4-m}}}
\,\left(1+{2\,A^2f^2\left(B^{-1}[K\,\phi]\right)^{-2(1-f)}\over
\kappa\,\tau}\right)^{-\frac{m}{2(4-m)}}, \label{gammaph}%
\end{equation}
for the case  $m \neq 4$.

From the definition of the dimensionless slow-roll parameter
$\varepsilon=-\frac{\dot{H}}{H^{2}}$,  then we find that
$
\varepsilon=\left(  {\frac{1-f}{Af}}\right)
{1\over (B^{-1}[K\,\phi])^{f}}.
$ In this form, and considering that
 the condition for inflation to occur is determined by $\varepsilon<$1 (or
equivalently $\ddot{a}>0$
), then  the scalar field during warm inflation satisfies the condition
$\phi>{1\over K}B\left[\left({\frac{1-f}{Af}}\right)^{1/f}\right]$.



On the other hand, from Eqs.(\ref{at}) and (\ref{fisol}),
 the number of e-folds $N$ between two  different values of the scalar field, denoted $\phi_{1}$ and
$\phi_{2}$, is

\begin{equation}
N=\int_{t_{1}}^{t_{2}}\,H\,dt=A\,\left(  t_{2}^{f}-t_{1}^{f}\right)
=A\,\left[(B^{-1}[K\,\phi_{2}])^{f}-(B^{-1}[K\,\phi_{1}])^{f} \right]. \label{N1}%
\end{equation}

Following Ref.\cite{Barrow3}, the  inflationary phase begins at
the earliest possible scenario. In this way,  the
scalar field $\phi_{1}$ takes the value
\begin{equation}
\phi_{1}={1\over K}B\left[\left({\frac{1-f}{Af}}\right)^{1/f}\right]\;. \label{al}%
\end{equation}

In the following, we will study the  scalar and tensor perturbations for our
brane-warm model during the weak dissipative regime.
For the case of the scalar perturbation, it could be stated as ${\mathcal{P}%
_{\mathcal{R}}}^{1/2}=\frac{H}{\dot{\phi}}\,\delta\phi$\cite{warm}.
It is well know that in
the warm inflation scenario, the scalar field fluctuations are predominantly
thermal rather than
quantum \cite{warm,62526}. In particular, for the weak
dissipation regime, the amplitude of the scalar field fluctuation $\delta\phi^{2}$ is given
by
$\delta\phi^{2}\simeq H\,T$
\cite{62526}.   In this way,  from
Eqs.(\ref{inf3}), (\ref{rh-1})
and (\ref{G1}), the power spectrum ${\mathcal{P}_{\mathcal{R}}}$,
results
\begin{equation}
{\mathcal{P}_{\mathcal{R}}}={3\sqrt{3\pi}\,\kappa\over 4}\, \left(  \frac{C_{\phi}}{6\kappa
C_{\gamma}}\right)  ^{{\frac{1}{4-m}}}\phi^{{\frac{1-m}{4-m}}}H^{{\frac
{11-3m}{4-m}}}(-\dot{H})^{-{\frac{(3-m)}{4-m}}}\left(1+{2 H^{2}\over \kappa\tau}
\right)^{{\frac{3-m}{2(4-m)}}}. \label{pd}%
\end{equation}

Now, combining  Eqs.(\ref{at}) and (\ref{fisol}), we find that  the power
spectrum as function of the field $\phi$ can be written as
\begin{equation}
{\mathcal{P}_{\mathcal{R}}}(\phi)=k_{1}\,\,\phi^{\,\frac{1-m}{4-m}}
\left(B^{-1}[K\,\phi] \right)  ^{\frac{2f(4-m)+m-5}{4-m}}
\left[ 1+\frac{2A^{2}f^{2}}
{\kappa\tau(B^{-1}[K\,\phi])^{2(1-f)}} \right]^{\frac{3-m}{2(4-m)}}, \label{pd}%
\end{equation}
where the constant $k_{1}$ is defined as
$
k_{1}={3\sqrt{3\pi}\,\kappa\over 4}\left(  \frac{C_{\phi}}{6\kappa
C_{\gamma}}\right)
^{{\frac{1}{4-m}}}(A\,f)^{{8-2m\over4-m}}%
\,(1-f)^{{\frac{m-3}{4-m}}}.
$

By other hand, the scalar power spectrum can be expressed in terms of the
number of $e$-folds $N$ as

\begin{equation}
{\mathcal{P}_{\mathcal{R}}}(N)=k_{2}\,\,(B[J(N)])^{\,\frac{1-m}{4-m}}
\left(J[N] \right)  ^{\frac{2f(4-m)+m-5}{4-m}}
\left[ 1+\frac{12A^{2}f^{2}}
{\kappa\tau (J[N])^{2(1-f)}} \right]^{\frac{3-m}{2(4-m)}},
\label{pdN}%
\end{equation}
where the quantities  $J(N)$ and $k_{2}$ are given by
$
J(N)=\left[{1+f(N-1)\over Af} \right]^{\frac{1}{f}},
$ and
$
k_{2}=k_{1}K^{-\frac{1-m}{4-m}}
$, respectively.

From the definition of the scalar spectral index $n_{s}$,  given by the relation
$n_{s}-1=\frac{d\ln \,{\mathcal{P}_{R}}}{d\ln k}$, and combining  Eqs.
(\ref{fisol}), and (\ref{pd}), the scalar spectral index can be written as

\begin{equation}
n_{s}=1-{5-m-2f(4-m)\over Af(4-m)(B^{-1}[K\,\phi])^f} +n_2 +n_3,
\label{nss1}%
\end{equation}
where $n_2$ and $n_3$ are defined as

$$
n_{2}={1-m \over 4-m}\sqrt{{2(1-f)\over 3 \kappa Af}}
{(B^{-1}[K\,\phi])^{-f/2}\over \phi}\left[ 1+\frac{2A^{2}f^{2}}
{\kappa\tau(B^{-1}[K\,\phi])^{2(1-f)}} \right]^{-1/4},
$$
and
$$
n_{3}=-{2Af(1-f)(3-m) \over \kappa \tau(4-m)}(B^{-1}[K\,\phi])^{-(2-f)}
\left[ 1+\frac{2A^{2}f^{2}}
{\kappa\tau(B^{-1}[K\,\phi])^{2(1-f)}} \right]^{-1}.
$$

Analogously as the scalar power spectrum, the scalar spectral index  can be expressed  in terms of
the number of $e$-folds. Considering Eqs.(\ref{N1}) and (\ref{al}),
yields
\begin{equation}
n_{s}=1-\frac{5-m-2f(4-m)}{(4-m)[1+f(N-1)]}+n_2 +n_3, \label{nswr}%
\end{equation}
where now
$$
n_{2}=K{1-m \over 4-m}\sqrt{{2(1-f)\over \kappa Af}}
{(J[N])^{-f/2}\over B[J(N)]}\left[1+\frac{12A^{2}f^{2}}
{\kappa\tau (J[N])^{2(1-f)}}\right]  ^{-1/4},
$$
and
$$
n_{3}=-{2Af(1-f)(3-m) \over \kappa \tau (4-m)}(J[N])^{-(2-f)}
\left[1+\frac{12A^{2}f^{2}} {\kappa\tau
(J[N])^{2(1-f)}}\right]^{-1}.
$$

It is well know that
 the generation of
tensor perturbations during the  inflationary scenario would produce
gravitational waves. The  power spectrum of the tensor perturbations is
more complicated in our model because in brane-world gravitons propagate into the
bulk. In this form, following Ref.\cite{pt}, the  power spectrum of the tensor perturbations
${\mathcal{P}}_{g}$ is given by
${\mathcal{P}}_{g}=24\kappa(H/2\pi)^{2}F^{2}(x)$. Here, $x$ is defined as
$
x=Hm_p\,\sqrt{3/(4\pi\tau)},
$
and the function $F(x)$ is given by
$$
F(x)=[\sqrt{1+x^2}-x^2\,\sinh^{-1}(1/x)]^{-1/2}.
$$
The function $F(x)$ is the correction to the standard General Relativity and arises  from the normalization of a graviton zero-mode \cite{pt}.

In this way, we may compute an important observational quantity, the tensor-to-scalar ratio $r$, defined as
$r={\mathcal{P}}_{g}/{\mathcal{P}_{\mathcal{R}}}$.
 Then considering Eq.(\ref{pd}) and ${\mathcal{P}}_{g}$ in terms  of the scalar field, the tensor-to-scalar ratio
 can be written as
\begin{equation}
r(\phi)={6\,\kappa\,A^2 f^2 \over \pi^2\,k_{1}}
\,\phi^{\,-\frac{1-m}{4-m}} \left(B^{-1}[K\,\phi]
\right)^{-\frac{3-m}{4-m}} \left[ 1+\frac{2A^{2}f^{2}}
{\kappa\tau(B^{-1}[K\,\phi])^{2(1-f)}}\right]^{-\frac{3-m}{2(4-m)}}F^{2}(\phi)
.\label{Rk}%
\end{equation}

Similarly, as before, the tensor-to-scalar ratio can be expressed  in terms of
the number of $e$-folds, yielding
\begin{equation}
r(N) = {6\,\kappa\,A^2 f^2 \over \pi^2\,k_{2}}
(B[J(N)])^{-\frac{1-m}{4-m}} \left(J[N] \right)^{-\frac{3-m}{4-m}}
\left[ 1+\frac{12A^{2}f^{2}} {\kappa\tau (J[N])^{2(1-f)}}
\right]^{-\frac{3-m}{2(4-m)}}F^{2}(N)
. \label{Rk11}%
\end{equation}

\begin{figure}[th]
{{\hspace{0cm}\includegraphics[width=3.5in,angle=0,clip=true]{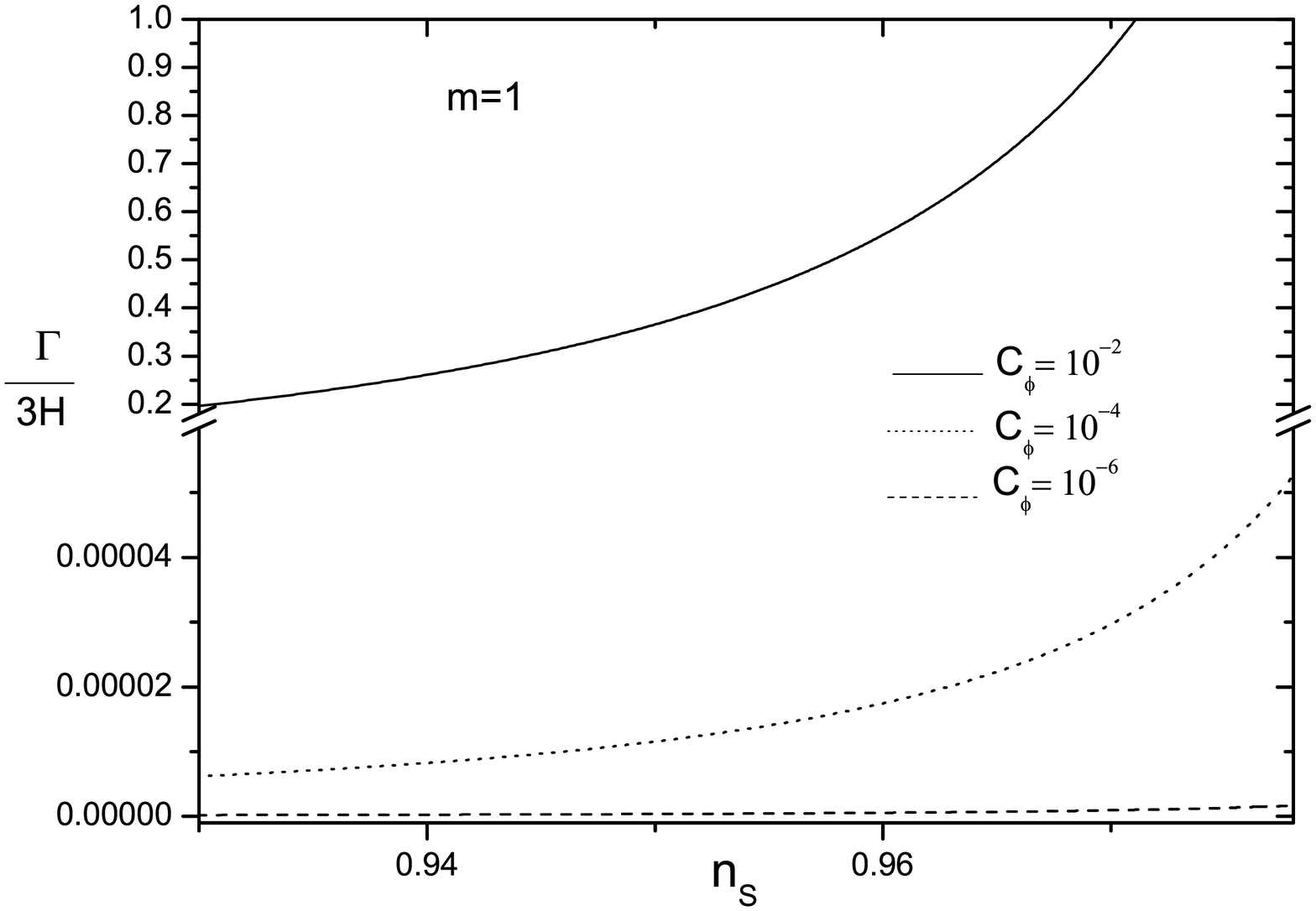}}}
{\includegraphics[width=3.5in,angle=0,clip=true]{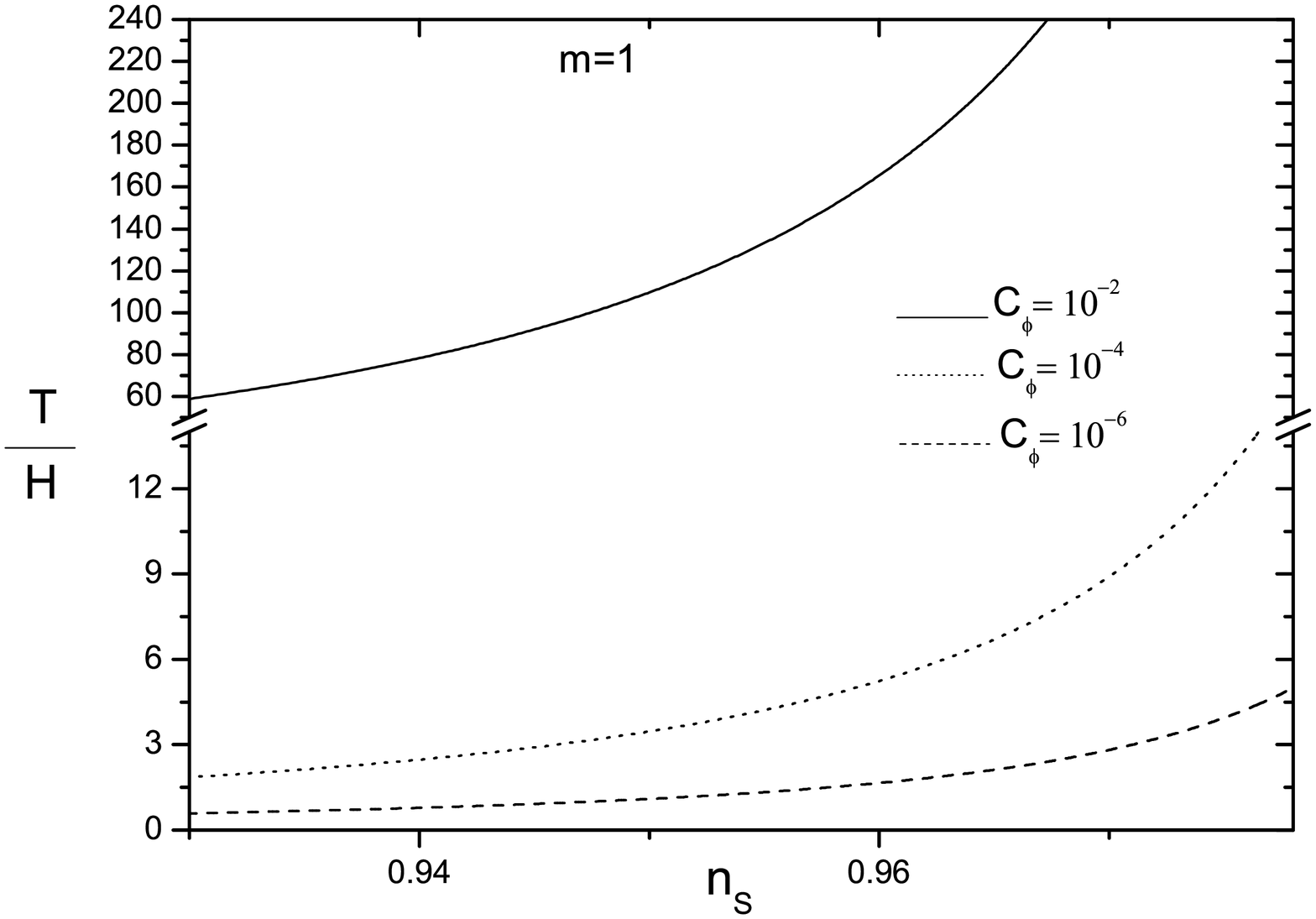}}
{\vspace{0. cm}\caption{ The evolution of the  ratio $\Gamma/3H$
versus the scalar spectrum index $n_s$ (upper panel) and the
evolution of the ratio $T/H$ versus  the scalar spectral index
$n_s$ (lower panel) in the weak dissipative regime for the special
case $m=1$ ($\Gamma\propto T$), for  different values of the
parameter $C_\phi$. In both panels, the solid, dotted, and dashed
lines correspond to the pairs ($A=0.19$, $f=0.31$), ($A=0.28$,
$f=0.32$) and($A=0.32$, $f=0.30$), respectively. Also, in these
plots we have used the values $C_\gamma=70$, $m_p=1$ and
$\tau=10^{-14}$.
 \label{fig1}}}
\end{figure}

\begin{figure}[th]
{{\hspace{-0.3cm}\includegraphics[width=4.1in,angle=0,clip=true]{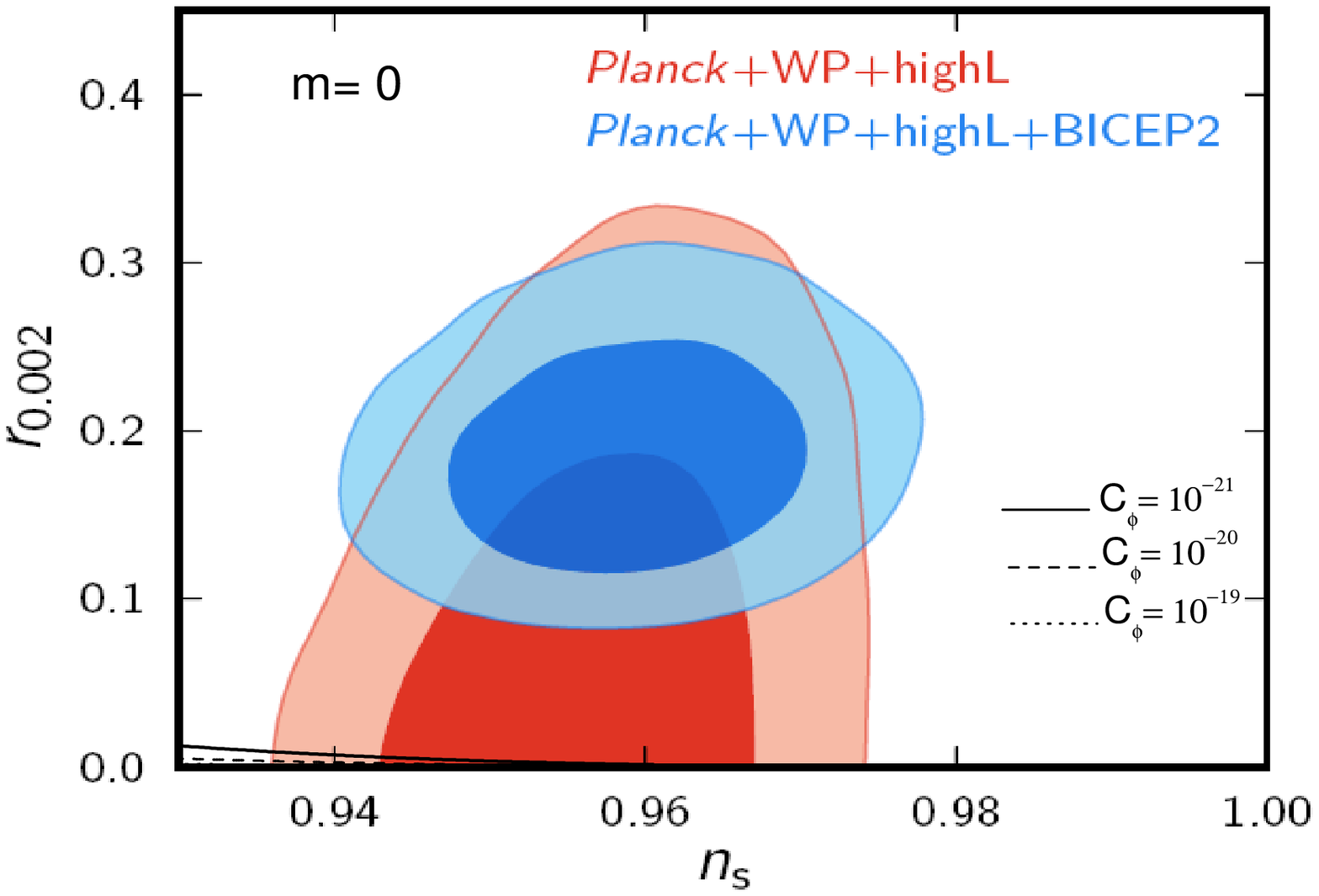}}}
{\vspace{-2.2cm}{{\hspace{-1.1cm}\includegraphics[width=3.8in,angle=0,clip=true]{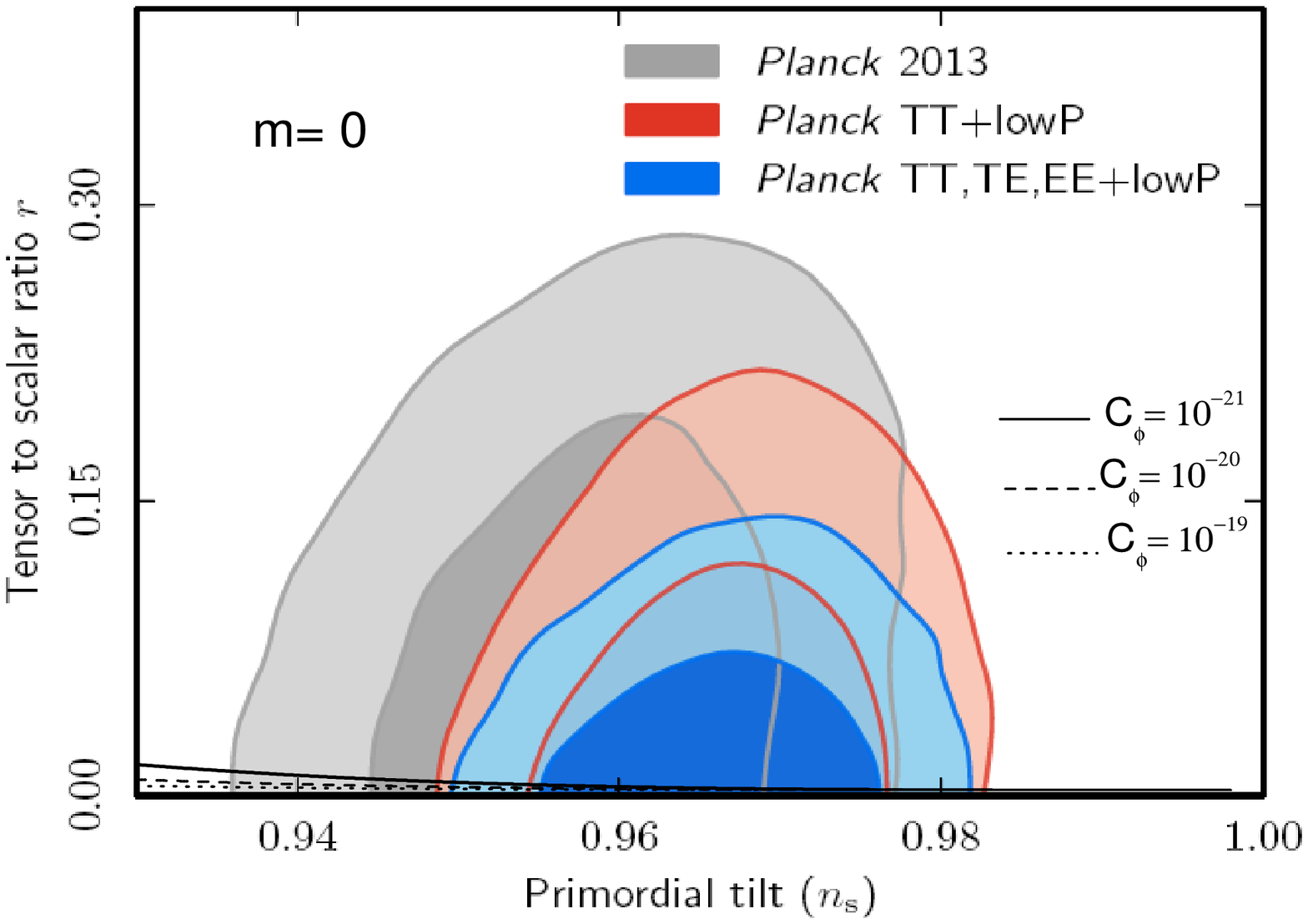}}}}
{\vspace{-2.0cm}\caption{ The upper and lower panels show the
evolution of the tensor-to-scalar ratio $r$ versus the scalar
spectral index $n_s$ in the weak dissipative regime for the
special case $m=0$ ($\Gamma\propto \phi$), for  different values
of the parameter $C_\phi$. In both panels, the solid, dotted, and
dashed lines correspond to the pairs ($A=0.47$, $f=0.29$),
($A=0.43$, $f=0.28$) and($A=0.39$, $f=0.29$), respectively. Also,
in both panels we have used the values $C_\gamma=70$, $m_p=1$,
$\tau=10^{-14}$, and the two-dimensional marginalized constraints
from Ref.\cite{BICEP} (upper panel) and the new data from Planck
2015 \cite{Planck2015}(lower panel).
 \label{fig12}}}
\end{figure}

In Fig.(\ref{fig1}) we show the evolution of the ratio $\Gamma/3H$
 versus the scalar spectral index (upper panel) and the evolution of
the ratio $T/H$ versus the the scalar spectral index (lower panel), in the
weak dissipative regime for the specific case in which the
dissipative coefficient is given by $\Gamma(\phi,T)=C_\phi\,T$,
i.e., $m=1$. In both panels, we have used three different values
of the parameter $C_\phi$. The upper panel shows the condition for
the weak dissipative regime in which $\Gamma<3H$. In the lower
panel we show the necessary condition for warm inflation scenario,
in which the temperature $T>H$. Combining Eqs.(\ref{at}),
(\ref{G1}) and (\ref{fisol}), we numerically find the ratio
$\Gamma/3H$ as a function of the scalar spectral index $n_s$. Analogously, from
Eqs.(\ref{at}) and (\ref{rh-1}), we numerically obtain the ratio between
the temperature of the thermal bath $T$ and the Hubble parameter $H$, i.e.,
$T/H$ in terms of the spectral
index $n_s$. In both plots, we consider the values $C_\gamma=70$, $m_p=1$
and $\tau=10^{-7}$. Also, numerically from Eqs.(\ref{pdN}) and (\ref{nswr}),
we obtain the values $A=0.19$ and $f=0.31$ corresponding to the value of the parameter
$C_\phi=10^{-2}$, for the values ${\mathcal{P}_{\mathcal{R}}}=2.43\times10^{-9}$,
 $n_s=0.96$, and the number of $e$-folds $N=60$.  Analogously, for the value $C_\phi=10^{-4}$, the
 values obtained for the parameters $A$ and $f$ are given by $A=0.28$ and $f=0.32$, respectively. Finally,for the
 value $C_\phi=10^{-6}$, we obtain the values $A=0.32$ and $f=0.30$. From the upper
 plot, we find an upper bound for the parameter $C_\phi$, from the condition of the
 weak dissipative regime, i.e., $\Gamma<3H$, for which $C_\phi\leq 10^{-2}$. From the lower panel
  we obtain a lower bound for $C_\phi$, considering the essential condition for
  warm inflation $T>H$, where $C_\phi\geq 10^{-6}$. In relation to the
  consistency relation $r=r(n_s)$,
  we find that the ratio $r\sim 0$ for this range of $C_\phi$,
 then   the case $m=1$ (or equivalently $\Gamma\propto T$) during the weak
 dissipative regime
 is disproved from BICEP2 experiment, because $r=0.2_{-0.05}^{+0.07}$  and
 further the ratio $r=0$ is discarded at 7.0$\sigma$. However,  the Planck data analysis
 obtained only an upper limit for the tensor-to-scalar ratio $r$, given by $r<0.11$, and then
 the range of $C_\phi$ is well corroborated from Planck satellite. In this form,
 for the specific case of $m=1$, the range of the parameter $C_\phi$ is given by
 $10^{-6}\leq C_\phi \leq 10^{-2}$.
 Also,
 we note that when we decrease the
value of the parameter $C_\phi< 10^{-6}$, the value of the tensor to scalar ratio
$r\simeq 0$. In particular, for the value $C_\phi=10^{-6}$,
we get $r\mid_{n_s=0.96}\simeq7\times10^{-4}$. It is interesting to note that the range
for the parameter $C_\phi$ for the case $\Gamma\propto T$ results from the conditions $\Gamma<3H$
and $T>H$.

In Fig.(\ref{fig12}) we show the tensor-to-scalar ratio $r$ versus
the scalar spectral index $n_s$, for the special case of $m=0$,
i.e., $\Gamma\propto \phi$ in the weak dissipative regime
($\Gamma<3H$). In the upper panel we show the two-dimensional
marginalized constraints for the tensor-to-scalar ratio $r=r(n_s)$
(at 68$\%$ and 95$\%$ levels of confidence), from the BICEP2
experiment data in connection with Planck satellite+ WP+
highL\cite{BICEP}. In the lower panel the new results from
Planck 2015\cite{Planck2015}. Here, the marginalized joint 68$\%$
and 95$\%$ CL regions for the spectral index $n_s$ and
$r_{0.002}$. From Eqs.(\ref{nswr}) and (\ref{Rk11}), we
numerically obtain the consistency relation $r=r(n_s)$ and, as
before, we consider three values of the parameter $C_\phi$. Again,
we take the values $C_\gamma=70$, $m_p=1$ and $\tau=10^{-14}$.
From these plots we observe that the tensor-to -scalar ratio
$r\sim 0$ for the specific case of $m=0$, and then this case is
disproved from BICEP2 experiment (upper panel), however is well
corroborated from Planck data and in special with the new data
(lower panel). We observe that the new results from Planck 2015
place more substantial limits on the consistency relation
$r=r(n_s)$ compared with the BICEP2 experiment.

In particular for the value of $C_\phi=10^{-21}$ (solid line in
the figure), we get that $r\mid_{n_s=0.96}\simeq 2\times 10^{-3}$.
From the essential condition for warm inflation $T>H$, we find a
lower limit for the parameter $C_\phi$ given by $C_\phi\geq
10^{-21}$ and, considering the condition of weak dissipative
regime, where $\Gamma<3H$, we obtain un upper bound for $C_\phi$,
and it corresponds to $C_\phi\leq 10^{-15}$ (figure not shown). In
this form, for the special case of $m=0$ (or equivalently
$\Gamma\propto \phi$), the range for the parameter $C_\phi$ is
given by $10^{-21}\leq C_\phi\leq 10^{-15}$.

For the cases $m=3$ and $m=-1$, we find that the tensor-to-scalar
ratio $r\sim 0$, then these cases are disproved from BICEP2 data,
nevertheless are well corroborated from Planck satellite and
Planck 2015 data. Considering the essential condition for warm
inflation $T>H$, we obtain a lower bound for the parameter
$C_\phi$; for the case $m=3$ the bound is $10^{22}\leq C_\phi$ and
for the case $m=-1$ it corresponds to $10^{-34}\leq C_\phi$. In
particular, for the value $C_\phi=10^{22}$ corresponds to
$\left(\frac{T}{H}\right)\mid_{n_s=0.96}\simeq 1.86$ when $m=3$,
and for  $C_\phi=10^{-34}$ corresponds to
$\left(\frac{T}{H}\right)\mid_{n_s=0.96}\simeq 1.57$ for the case
$m=-1$. From the condition of weak dissipative regime $\Gamma<3H$,
we find an upper bound for $C_\phi$; for the case $m=3$ the upper
bound is $C_\phi\leq 10^{23}$, and for the specific case $m=-1$,
results $C_\phi\leq 10^{-28}$. In particular, for the value
$C_\phi=10^{23}$ corresponds to
$\left(\frac{\Gamma}{3H}\right)\mid_{n_s=0.96}\simeq 0.44$ when
$m=3$, and for  $C_\phi=10^{-28}$ corresponds to
$\left(\frac{\Gamma}{3H}\right)\mid_{n_s=0.96}\simeq 0.18$ for the
case $m=-1$. In this form, from the conditions $T>H$ and
$\Gamma<3H$, the ranges for the parameter $C_\phi$, are given by:
$10^{22}\leq C_\phi\leq 10^{23}$ for the case $m=3$, and
$10^{-34}\leq C_\phi\leq 10^{-28}$ for the case $m=-1$.

\subsection{ The strong dissipative regime.\label{subsection2}}

Now we consider the case of strong dissipative regime $R>1$
(or equivalently  $\Gamma>3H$), together with the scalar factor $a(t)$ of intermediate inflation,
see Eq.(\ref{at}). Considering   Eqs. (\ref{inf3}) and (\ref{G1}),  we find the
solution for the scalar field $\phi(t)$.
In particular, we must to analyze our solution for two separate cases, namely  $m=3$ and $m\neq3$.
For the special  case of $m=3$, the solution for the scalar field results
\begin{equation}
\phi(t)-\phi_0=\frac{\tilde{B}[t]}{\tilde{K}}, \label{phisf}%
\end{equation}
as before, the value of $\phi(t=0)=\phi_{0}$ is an integration constant and  $\tilde{K}$ is a constant
given by
$$
\tilde{K}\equiv\left(\frac{4C_{\phi}}{\tau}\right)^{1/2}\left(\frac{1}{2\kappa C_{\gamma}}\right)^{3/8}\left[\left(\frac{\kappa \tau}{2}\right)^{\frac{4-3(f-2)}{2(1-f)}}\left(\frac{(1-f)^{1-f}}{Af}\right)^{7}\right]^{1/8(1-f)},\label{Ktilde}
$$
and $\tilde{B}[t]$ is a new function that  is defined as
\begin{equation}
\tilde{B}[t]\equiv B\left[-\frac{\kappa \tau}{2(Af)^2}t^{2(1-f)};\,\frac{4f+3}{16(1-f)},\,
\frac{15}{16}\right],\label{Betatilde}
\end{equation}
and corresponds to the incomplete beta function\cite{Libro}.

On the other hand, the solution of the scalar field for the special case $m\neq3$
yields
\begin{equation}
\varphi(t)-\varphi_0=\frac{\tilde{B}_{m}[t]}{\tilde{K}_{m}}, \label{phisfm}%
\end{equation}
here the  scalar field $\phi(t)$ is redefined as
$\varphi(t)=\frac{2}{3-m}\phi(t)^{\frac{2}{3-m}}$
 and as before $\varphi_0$ is an integration constant, that can be assumed $\varphi_0=0$.
The quantity $\tilde{K}_m$ is a new constant given by
$\tilde{K}_m\equiv\left(\frac{4C_{\phi}}{\tau}\right)^{1/2}\left(\frac{1}
{2\kappa C_{\gamma}}\right)^{m/8}\left[\left(\frac{\kappa \tau}{2}\right)^
{\frac{4-m(f-2)}{2(1-f)}}\left(\frac{(1-f)^{1-f}}{Af}\right)^{4+m}\right]^{1/8(1-f)}$, and the
new function $\tilde{B}_m[t]$ for the special case in which $m\neq3$
is defined as
\begin{equation}
\tilde{B}_m[t]\equiv B\left[-\frac{\kappa \tau}{2(Af)^2}t^{2(1-f)};
\,\frac{4f+m}{16(1-f)},\,\frac{12+m}{16}\right].\label{Betatildem}
\end{equation}

From Eqs.(\ref{at}), (\ref{phisf}) and (\ref{phisfm}), we find that the Hubble parameter $H=H(\phi)$
yields
\begin{equation}
H(\phi)=\frac{Af}{(\tilde{B}^{-1}[\tilde{K}\ln \phi])^{1-f}},\,\,\;\text{for}\,m=3,
\end{equation}
and for the special case of $m\neq 3$ we have
\begin{equation}
H(\phi)=\frac{Af}{(\tilde{B}_m^{-1}[\tilde{K}_m\varphi])^{1-f}}.
\end{equation}

By considering  Eq.(\ref{pot}) ,
the scalar potential under the slow-roll approximation for both values of $m$
is given by
\begin{equation}
V(\phi)= \tau\left(-1+\left[1+\frac{2A^2f^2}{\kappa
\tau (\tilde{B}^{-1}[\tilde{K}\ln\phi])^{2(1-f)}}\right]^{1/2}\right),\label{potm3}
\end{equation}
for the special case  $m=3$, and
\begin{equation}
V(\phi)=\tau\left(-1+\left[1+\frac{2A^2f^2}{\kappa
\tau (\tilde{B}_m^{-1}[\tilde{K}_m\varphi])^{2(1-f)}}\right]^{1/2}\right),\label{potmd3}
\end{equation}
for the value of $m\neq3$.

 The dissipative coefficient $\Gamma$ in terms of the scalar field, can be obtained combining
  Eqs. (\ref{G1}), (\ref{phisf}),
  and (\ref{phisfm}) to give
  \begin{equation}
\Gamma(\phi)=\delta \phi^{-2}(\tilde{B}^{-1}[\tilde{K}\ln\phi])^{-\frac{3(2-f)}{4}}\left[1+\frac{2H^2}{\kappa \tau}\right]^{-3/8}, \label{gammas3}%
\end{equation}
for the case $m=3$, in which  the constant $\delta$ is defined as
 $\delta=C_{\phi}\left[\frac{Af(1-f)}{2\kappa C_{\gamma}}\right]^{3/4}$.  For the value of $m\neq3$
 we find
\begin{equation}
\Gamma(\phi)=\delta_m \phi^{1-m}(\tilde{B}_m^{-1}[\tilde{K}_m\varphi])^{-\frac{m(2-f)}{4}}\left[1+\frac{2H^2}{\kappa \tau}\right]^{-m/8}, \label{gammasm}%
\end{equation}
where $\delta_m$ is a  constant  and is given by $\delta_m=C_{\phi}\left[\frac{Af(1-f)}{2\kappa C_{\gamma}}\right]^{m/4}$.

Analogous to the case of  the weak dissipative regime,
 the dimensionless slow-roll parameter $\varepsilon$ is given by $
\varepsilon=-\frac{\dot{H}}{H^{2}}=\frac{1-f}{Af(\tilde{B}^{-1}[\tilde{K}\ln\phi])
^f}$, for  $m=3$ and  for the value of
$m\neq3$ we find $\varepsilon=\frac{1-f}{Af(\tilde{B}_m^{-1}[\tilde{K}_
m\varphi])^f}$. Again, if
 $\ddot{a}>0$, then
the scalar field $\phi>\exp[\frac{1}{\tilde{K}}\tilde{B}[(\frac{1-f}{Af})^{1/f}]]$,
 for the special  case $m=3$, and
for the case $m\neq3$ the new scalar field satisfied
$\varphi>\frac{1}{\tilde{K}_m}\tilde{B}_m[(\frac{1-f}{Af})^{1/f}]$. Analogously
as before,
 the inflationary scenario begins ($\varepsilon=1$) when the scalar field takes
 the value
$\phi_1=\exp[\frac{1}{\tilde{K}}\tilde{B}[(\frac{1-f}{Af})^{1/f}]]$, for  $m=3$,
and $\varphi_1=\frac{1}{\tilde{K}_m}\tilde{B}_m[(\frac{1-f}{Af})^{1/f}]$ for the
special case of
 $m\neq3$.

 The number of e-folds $N$ in this regime can be write
  using  Eqs.(\ref{at}), (\ref{phisf}), and (\ref{phisfm}), to give
\begin{equation}
N=\int_{t_{1}}^{t_{2}}\,H\,dt=A[(\tilde{B}^{-1}[\tilde{K}\ln\phi_2])^f-(\tilde{B}^{-1}[\tilde{K}\ln\phi_1])^f],\label{Nsr3}%
\end{equation}
for the case of $m=3$, and for the special case of  $m\neq3$ we have
\begin{equation}
N=A[(\tilde{B}_m^{-1}[\tilde{K}_m\varphi_2])^f-(\tilde{B}_
m^{-1}[\tilde{K}_m\varphi_1])^f].\label{Nsrm}%
\end{equation}

Analogous to the case of the weak dissipative regime, now  we will analyze
the cosmological perturbations in which $R=\Gamma/3H>1$ (strong dissipative regime).
For the strong dissipative regime, the fluctuations $\delta\phi^2$ is given by
 $\delta\phi ^{2}\simeq\frac{k_{F}T}{2\pi^{2}}$, see Ref.\cite{warm} ,
where $k_{F}$ corresponds to the wave-number and  it is determined as  $k_{F}=\sqrt{\Gamma
H}=H\sqrt{3R}>H$. In this way,  the power spectrum
of the scalar perturbation in this regime from Eqs.(\ref{at}),
(\ref{rh-1}) and (\ref{G1}),  can be written as
\begin{equation}
P_{\mathcal{R}}\simeq\frac{H^{\frac{5}{2}}\Gamma^{\frac{1}{2}}T}{2\pi^{2}%
\dot{\phi}^{2}}=\frac{\kappa}{4\pi^2}C_{\phi}^{3/2}\phi^{\frac{3(1-m)}{2})}H^{3/2}(-\dot{H})^{\frac{3m-6}{8}}\left[1+\frac{2H^2}{\kappa \tau}\right]^{-\frac{(3m-6)}{16}}.\label{Prsr}%
\end{equation}

Analogously as before, we can find the power spectrum $\mathcal{P}_{\mathcal{R}}$
 in terms of the
scalar field $\phi$ for both values of the parameter $m$. In this form,
considering Eqs. (\ref{at}), (\ref{phisf}), (\ref{phisfm}) and
Eq. (\ref{Prsr}) yields
\begin{equation}
\mathcal{P}_{\mathcal{R}}=k(\tilde{B}^{-1}[\tilde{K}\ln\phi])^
{\frac{3(5f-6)}{8}}\phi^{-3}\left[1+\frac{2(Af)^2(\tilde{B}^{-1}
[\tilde{K}\ln\phi])^{-2(1-f)}}{\kappa \tau}\right]^{-3/16},\label{Prf3}
\end{equation}
for the case of $m=3$. Here
 the constant $k$ is given by $k=\frac{k}{4\pi^2}C_{\phi}^{3/2}
 \left(\frac{1}{2\kappa C_{\gamma}}\right)^{11/8}(Af)^{15/8}(1-f)^{3/8}$.

 Now, the spectrum of the scalar perturbation
 for the special case of $m\neq3$,  results
\begin{equation}
\mathcal{P}_{\mathcal{R}}=k_m(\tilde{B}_m^{-1}[\tilde{K}_\varphi])^{\frac{3[f(2+m)-2m]}{8}}\phi^{\frac{3(1-m)}{2}}\left[1+\frac{2(Af)^2(\tilde{B}_m^{-1}[\tilde{K}_m\varphi])^{-2(1-f)}}{\kappa \tau}\right]^{-\frac{(3m-6)}{16}},\label{Prfm}
\end{equation}
where $k_m$ is a constant given by $k_m=\frac{k}{4\pi^2}C_{\phi}^{3/2}\left(\frac{1}{2\kappa C_{\gamma}}\right)^{\frac{3m+2}{8}}(Af)^{\frac{3m+6}{8}}(1-f)^{\frac{3m-6}{8}}$.

Also, the scalar power spectrum can be rewritten  in terms of the
number of $e$-folds $N$.
 From Eqs.(\ref{Nsr3}) and (\ref{Nsrm}),  the power spectrum
  $\mathcal{P}_{\mathcal{R}}$ becomes
\begin{equation}
\mathcal{P}_{\mathcal{R}}=k(J[N])^{\frac{3(5f-6)}{8}}\exp\left
(-\frac{\tilde{B}[J[N]]}{\tilde{K}}\right)\left[1+\frac{2(Af)^2(J[N])^{-2(1-f)}}
{\kappa \tau}\right]^{-3/16},\label{Prf3N}
\end{equation}
for the specific case of $m=3$, and for the case $m\neq3$ we have
\begin{equation}
\mathcal{P}_{\mathcal{R}}=\tilde{\gamma}_m(J[N])^{\frac{3[f(2+m)-2m]}{8}}(\tilde{B}_m[J[N]])^{\frac{3(1-m)}{3-m}}\left[1+\frac{2(Af)^2(J[N)^{-2(1-f)}}{\kappa \tau}\right]^{-\frac{(3m-6)}{16}},\label{PrfmN}
\end{equation}
where the constant $\tilde{\gamma}_m$  in the above equation is given by $\tilde{\gamma}_m=k_m\left(\frac{2\tilde{K}_m}{3-m}\right)^{-\frac{3(1-m)}{3-m}}$.

For this regime the scalar spectral index $n_s$ for the specific case of $m=3$
from Eqs. (\ref{Prf3}) and (\ref{Prfm}) results
\begin{equation}
n_s=1+\frac{3(5f-6)}{8Af}(\tilde{B}^{-1}[\tilde{K}\ln\phi])^{-f}+
n_1+n_2.\label{nsf3}
\end{equation}
Here the new functions $n_1$ and $n_2$ are given by $n_1=-3\left(\frac{2}{\kappa}\right)^{1/2}\frac{1}{C_{\phi}^{1/2}}\left(\frac{1}{2\kappa C_{\gamma}}\right)^{-3/8}(Af)^{-3/8}\\\times (1-f)^{1/8}\left[1+\frac{2(Af)^2(\tilde{B}^{-1}[\tilde{K}\ln\phi])^{-2(1-f)}}{\kappa \tau}\right]^{-1/16}$ and
$n_2=\frac{3Af(1-f)}{4\kappa \tau}(\tilde{B}^{-1}[\tilde{K}\ln\phi])
^{f-2}\\\times\left[1+\frac{2(Af)^2(\tilde{B}^{-1}[\tilde{K}\ln\phi])^{-2(1-f)}
}{\kappa \tau}\right]^{-1}$, respectively. By other hand,  the scalar spectral index for the case
 $m\neq3$ becomes

\begin{equation}
n_s=1+\frac{3[f(m+2)-2m]}{8Af}(\tilde{B}_m^{-1}[\tilde{K}_m\varphi])^{-f}+n_{1_{m}}+n_{2_{m}},\label{nsfm}
\end{equation}
where  now the new functions $n_{1_{m}}$ and $n_{2_{m}}$ are defined by $n_{1_{m}}=\frac{3(1-m)}{2Af}\left(\frac{2}{\kappa}\right)^{1/2}\left(\frac{1}{2\kappa C_{\gamma}}\right)^{-m/8}\\\times\frac{1}{C_{\phi}^{1/2}}(Af)^{\frac{8-m}{8}}(1-f)^{\frac{4-m}{8}}(\tilde{B}_m^{-1}[\tilde{K}_m\varphi])^{-\frac{[4+m(f-2)]}{8}}\phi^{\frac{m-3}{2}}
\left[1+\frac{2(Af)^2(\tilde{B}_m^{-1}[\tilde{K}_m\varphi])^{-2(1-f)}}{\kappa \tau}\right]^{\frac{m-4}{16}}$
and $n_{2_{m}}=\frac{(3m-6)}{4\kappa \tau}(Af)(1-f)(\tilde{B}_m^{-1}[\tilde{K}_m\varphi])^{f-2}\left[1+\frac{2(Af)^2(\tilde{B}_m^{-1}[\tilde{K}_m\varphi])^{-2(1-f)}}{\kappa \tau}\right]^{-1}$.

Analogously as before, from Eqs.(\ref{Nsr3}) and (\ref{Nsrm}), the
scalar spectral index $n_s$ can be rewritten  in terms of the number
of $e$-folds as
\begin{equation}
n_s=1+\frac{3(5f-6)}{8Af}(J[N])^{-f}+n_1+n_2,\label{nsf3N}
\end{equation}
for the specific case of $m=3$. Here the quantities $n_1(J[N])$ and $n_2(J[N])$
 are defined as $n_1(J[N])=-3\left(\frac{2}{\kappa}\right)^{1/2}\frac{1}{C_{\phi}^{1/2}}\left(\frac{1}{2\kappa C_{\gamma}}\right)^{-3/8}(Af)^{-3/8} (1-f)^{1/8}\left[1+\frac{2(Af)^2(J[N])^{-2(1-f)}}{\kappa \tau}\right]^{-1/16}$
and $n_2(J[N])=\frac{3Af(1-f)}{4\kappa \tau}(J[N])^{f-2}\left[1+
\frac{2(Af)^2(J[N])^{-2(1-f)}}{\kappa \tau}\right]^{-1}$, respectively.
Analogously, the scalar spectral index for the value of $m\neq3$ yields
\begin{equation}
n_s=1+\frac{3[f(m+2)-2m]}{8Af}(J[N])^{-f}+n_{1_{m}}+n_{2_{m}},\label{nsfmN}
\end{equation}
where $n_{1_{m}}(J[N])=\frac{6(1-m)}{3-m}\left(\frac{\kappa}{2}\right)^{\frac{[f(16-m)+2(m-6)]}{16(1-f)}}(Af)^{-\frac{[4+m(2-f)]}{8(1-f)}}\left[1+\frac{2(Af)^2(J[N])^{-2(1-f)}}{\kappa \tau}\right]^{\frac{m-4}{16}}\\\times(J[N])^{-\frac{[4+m(f-2)]}{8(1-f)}}(\tilde{B}_m[J[N]])^{-1}$
and
$n_{2_{m}}(J[N])=\frac{(3m-6)}{4\kappa \tau}(Af)(1-f)(J[N])^{f-2}\\\times\left[1+\frac{2(Af)^2(J[N])^{-2(1-f)}}{\kappa \tau}\right]^{-1}$.

Analogous to the case of the weak dissipative regime,
 the tensor-to-scalar-ratio $r$ for the specific case $m=3$ can be expressed
in terms of the scalar field to give
\begin{equation}
r=\frac{6\kappa}{\pi^2 k}(Af)^2(\tilde{B}^{-1}[\tilde{K}\ln\phi])^{\frac{(f+2)}{8}}\phi^3\left[1+\frac{2(Af)^2(\tilde{B}^{-1}[\tilde{K}\ln\phi])^{-2(1-f)}}{\kappa \tau}\right]^{\frac{3}{16}}F^2(\phi),\label{rs3f}
\end{equation}
and  the tensor-to-scalar-ratio for  the case $m\neq3$ yields
$$
r=\frac{6\kappa}{\pi^2 k_m}(Af)^2(\tilde{B}_m^{-1}[\tilde{K}_m\varphi])^{
\frac{1}{8}[6m+f(10-3m)-16]}\phi^{\frac{3}{2}(m-1)}\,\,\;\times
$$
\begin{equation}
\left[1+\frac{2(Af)^2(\tilde{B}_m^{-1}[\tilde{K}_m\varphi])^{-2(1-f)}}{\kappa \tau}\right]^{\frac{(3m-6)}{16}}F^2(\phi).\label{rsmf}
\end{equation}

Also, we obtain the tensor-to-scalar ratio in terms of the
 number of $e$-folds.  In this form, combining Eqs.(\ref{Nsr3}) and
  (\ref{rs3f}) we get
\begin{equation}
r=\frac{6\kappa}{\pi^2 k}(Af)^2(J[N])^{\frac{(f+2)}{8}}\exp\left[3\frac{\tilde{B}[J[N]]}{\tilde{K}}\right]\left[1+\frac{2(Af)
^2(J[N])^{-2(1-f)}}{\kappa \tau}\right]^{\frac{3}{16}}F^2(N),\label{rs3N}
\end{equation}
for the specific case of $m=3$. From Eqs.(\ref{Nsrm}) and (\ref{rsmf}),
the tensor to scalar ratio becomes
$$
r=\frac{6\kappa}{\pi^2 k_m}(Af)^2(J[N])^{\frac{1}{8}[6m+f(10-3m)-16]}
\left(\frac{3-m}{2}\frac{\tilde{B}_m[J[N]]}{\tilde{K}_m}\right)^{\frac{3(m-1)}
{3-m}}\,\,\times
$$
\begin{equation}
\left[1+\frac{2(Af)^2(J[N])^{-2(1-f)}}{\kappa \tau}\right]^
{\frac{(3m-6)}{16}}F^2(N),\label{rsmN}
\end{equation}
for the case of $m\neq3$.

\begin{figure}[th]
{{\hspace{-3.5cm}\includegraphics[width=4.2in,angle=0,clip=true]{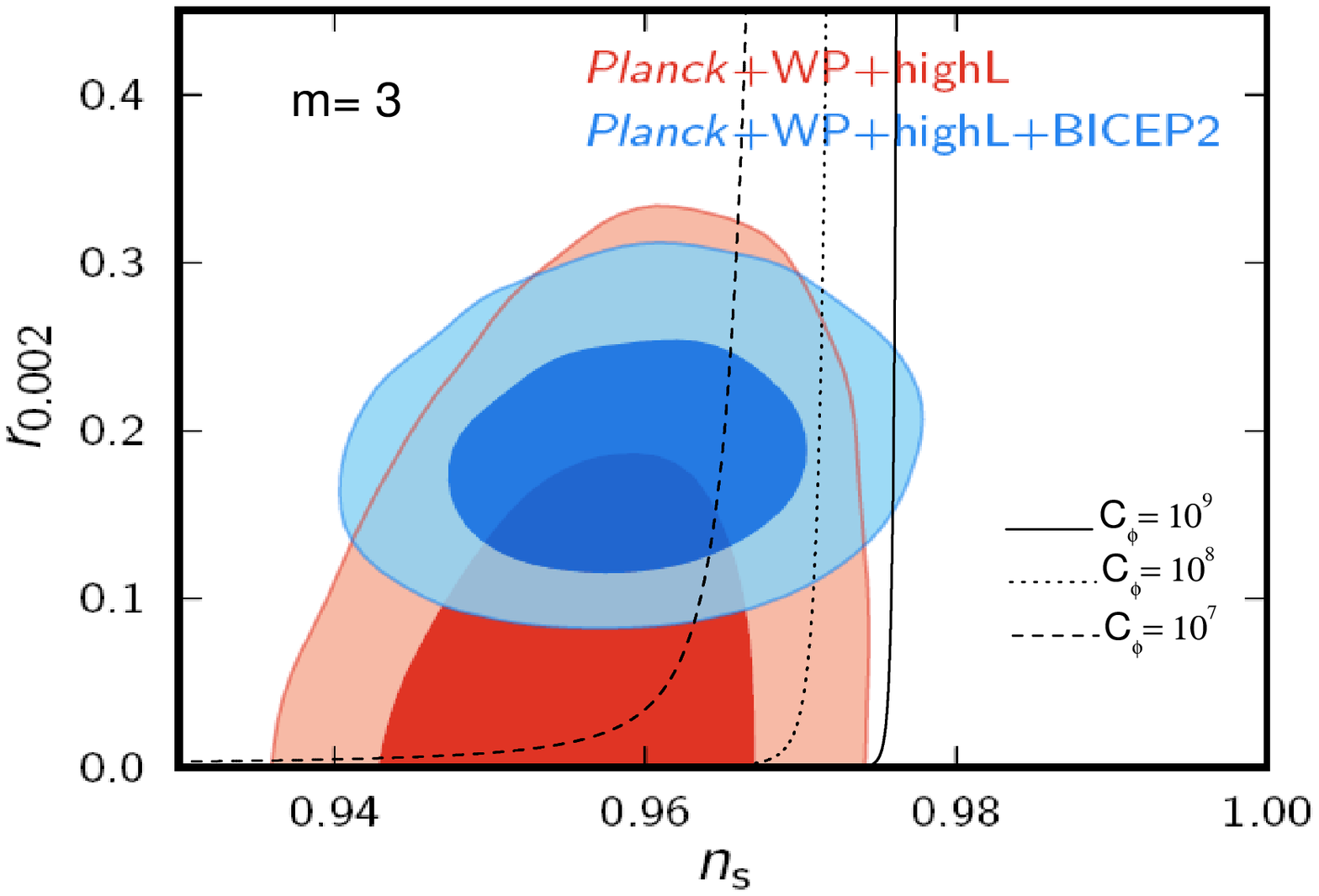}}}
{\vspace{-1.9 cm}
{{\hspace{+3.cm}\includegraphics[width=3.0in,angle=0,clip=true]{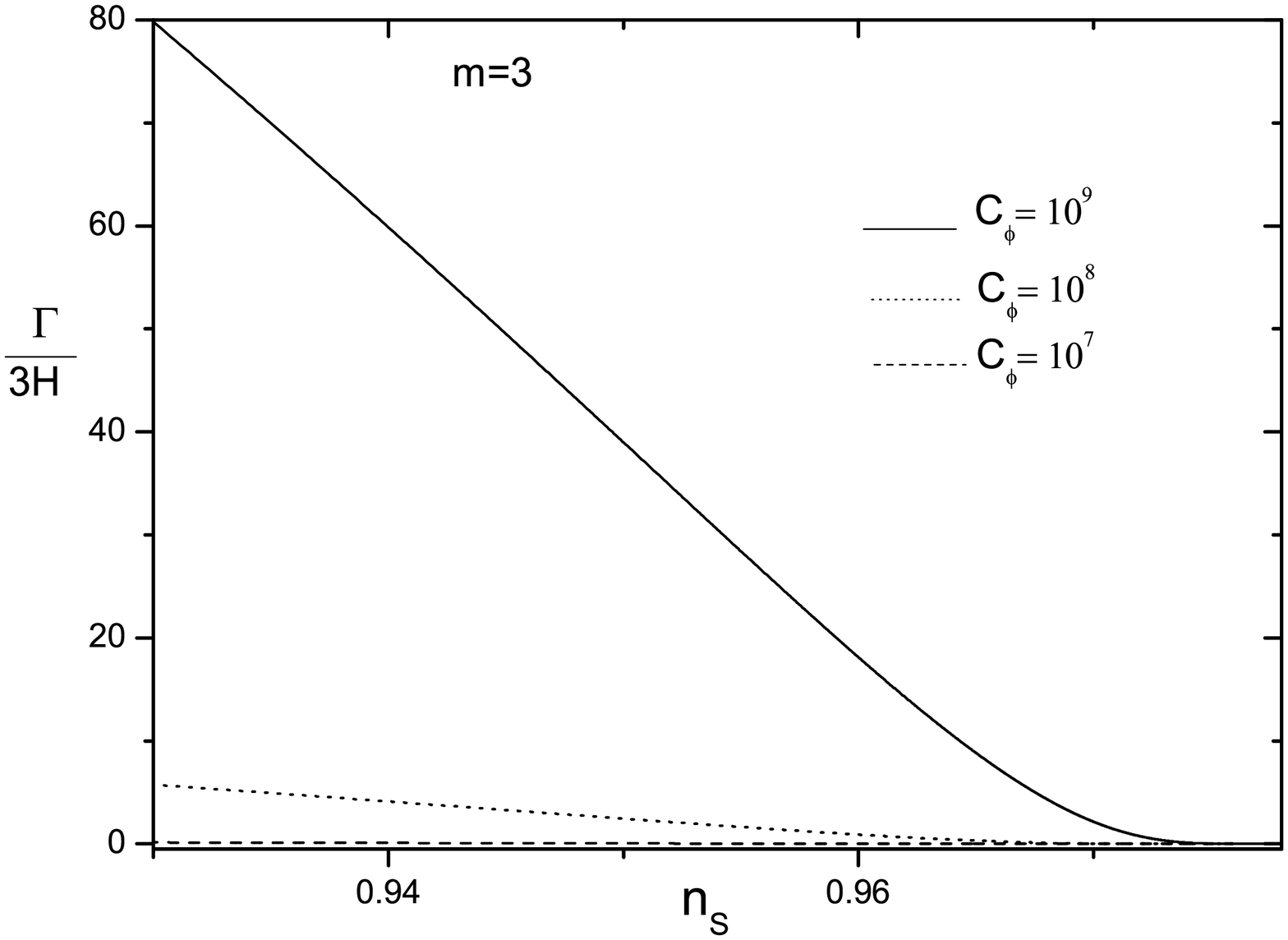}}}}
{\vspace{-0.5 cm}\caption{ The evolution of the tensor-to-scalar ratio
$r$ (upper panel) and the evolution of the
 ratio $\Gamma/3H$ versus the scalar spectral
index $n_s$ (lower panel) in the strong dissipative regime, for the
special case $m=3$ ($\Gamma\propto T^3/\phi^2$)  for three
different values of the parameter $C_\phi$.  In both  panels, the
solid, dotted, and dashed lines correspond to the pairs
($A=1.32\times10^{-5}$, $f=0.75$), ($A=7.68\times10^{-6}$,
$f=0.87$) and ($A=5.40\times10^{-6}$, $f=0.97$).
In both panels we have used the values $C_\gamma=70$,
$m_p=1$, $\tau=10^{-14}$, and the two-dimensional marginalized
constraints from Ref.\cite{BICEP}.
 \label{fig2}}}
\end{figure}

In Fig.(\ref{fig2}) we show the evolution of the tensor-to-scalar
ratio $r$ and the ratio $\Gamma/3H$ on the spectral index $n_s$ in
the strong dissipative regime, for the special case in which
$\Gamma=C_\phi\,T^3/\phi^2$, i.e., $m=3$. In both panels we have
used three different values of the parameter $C_\phi$. In the
upper panel, as before, we show the two-dimensional constraints in
the $n_s$-$r$ plane from BICEP2 and Planck data ($68\%$ and $95\%$
levels of confidence)\cite{BICEP}. In the lower panel we show the
evolution of the ratio $\Gamma/3H$ during the warm inflation scenario,
and  we corroborate that our model satisfies
the strong dissipative regime, i.e., $\Gamma>3H$. In order to write down
the consistency relation $r=r(n_s)$, we numerically find from
Eqs.(\ref{nsf3N}) and (\ref{rs3N}),   the
relation $r=r(n_s)$ (upper panel). Similarly, from
Eqs.(\ref{gammas3}), (\ref{Nsrm}) and (\ref{nsf3N}) we obtain the ratio
$\Gamma/3H$ as a function of the spectral index $n_s$ (lower
panel). In these plots we consider the values of $C_\gamma=70$, $m_p=1$ and
$\tau=10^{-14}$. Also, numerically from Eqs.(\ref{Prf3N}) and (\ref{nsf3N})
for the special case of $m=3$, we obtain  the
values $A=1.32\times10^{-5}$ and $f=0.75$ corresponding to the value $C_\phi=10^9$, for
$P_{\mathcal{R}}=2.43\times10^{-9}$, $n_s=0.96$  and
$N=60$. Analogously, for the parameter $C_\phi=10^8$, we numerically find the
the values $A=7.68\times10^{-6}$ and
$f=0.87$. Finally, for the value $C_\phi=10^7$ we obtain $A=5.40\times10^{-6}$ and
$f=0.97$. From the upper panel we obtain that the range for the parameter $C_{\phi}$ is
$10^7\leq C_\phi\leq 10^9$, which is well supported from observational data.
However, from the lower panel we observe that for values of $C_\phi\leq 10^8$ the model  is
disapproved from the condition the strong dissipative regime, since
the ratio $\Gamma/3H<1$. Here, we note that from the condition of the strong dissipative
 regime, i.e., $\Gamma>3H$, we have obtained a lower bound for the parameter
 $C_\phi$.
In this form, for the value $m=3$
( in which $\Gamma\propto T^3/\phi^2$)  the range for the parameter $C_\phi$ is
given by $10^8\leq C_\phi\leq 10^9$, which is well supported from observational
data together with the conditions  of the strong dissipative regime  $\Gamma>3H$ and $T>H$.

During the strong dissipative regime
for the specific case of $m=1$ ($\Gamma\propto T$) , we obtain
 that for the value of
the parameter $C_\phi\geq 10^{-1}$ the model is well corroborated from the
condition
$\Gamma>3H$ and the necessary condition for warm inflation $T>H$
 (figure not shown).  For the tensor-to-scalar ratio, we find that the ratio $r\sim 0$ for this lower bound,
 then the case $m=1$
 is disproved from BICEP2 experiment, since the ratio $r=0$ is discarded at 7.0$\sigma$. However,  from the Planck data,
 the value $C_\phi\geq 10^{-1}$ is well supported. Also,
 we note that when we increase the
value of the parameter $C_\phi> 10^{-1}$, the value of the tenso-to-scalar ratio
$r\simeq 0$.

For the cases $m=0$ ($\Gamma\propto \phi$) and $m=-1$
($\Gamma\propto \phi^2/T$), we find that these models in the strong dissipative models are
disproved from BICEP2 and Planck data, because the scalar spectral index $n_s>1$,
and then these models do not work.

\section{Conclusions \label{conclu}}

In this paper we have studied warm-intermediate inflation in the context of
Randall-Sundrum II brane-world cosmological model. Considering the slow-roll approximation during the weak
and strong regime, we have
obtained analytical solutions of the full effective Friedmann equation for a flat Universe
 in this brane-world model. Here we have considered a standard scalar field $\phi$
  together with a general form of the dissipative coefficient
  $\Gamma\propto\,T^m/\phi^{m-1}$. In special, we analyzed the values $m=3$,
  $m=1$, $m=0$ and $m=-1$, that can be have found in the literature
   for this dissipation coefficient.
  Studying the weak and strong dissipative regimes, we have obtained
  analytical expressions for the appropriate Hubble parameter, effective
  potential, scalar power spectrum, scalar spectral index and the tensor-to-scalar ratio.
    During both regimes
   we have studied the slow roll
   analysis  and we compared  with
   the two-dimensional marginalized  constraints
   (68$\%$ and 95$\%$ C.L.) $r=r(n_s)$ plane from observational data.
Also, we have obtained a constraint for the parameter $C_\phi$
 (see Eq.\ref{G})  from BICEP2 and Planck 2015 data together with the essential condition
  for warm inflation $T>H$ and the condition from
   the weak $\Gamma<3H$ (or strong $\Gamma>3H$) regime.

For all the models (different values of the parameter $m$)
 in the weak dissipative regime, we have  found a lower bound for
  the parameter $C_\phi$,
 from the essential
 condition for warm inflation, in which the temperature of the thermal bath
 $T>H$. Also, we have obtained  an upper bound for $C_\phi$, from the condition
 $\Gamma<3H$, i.e., the weak dissipative regime. Additionally, we have
  observed that
 the consistency relation $r=r(n_s)\sim 0$,
   in the weak dissipative scenario, and the models
     are disproved from BICEP2,
    but are well corroborated from Planck satellite, since $r<0.11$.

   For the strong dissipative scenario,  we have found that the range for the parameter
   $C_\phi$ is given by $10^8\leq C_\phi\leq 10^9$ in the specific case of $m=3$ i.e.,
   $\Gamma\propto T^3/\phi^2$.  Here, we have found an upper
    bound from BICEP2-Planck $r-n_s$ plane and a lower bound from the condition
    of the dissipative regime in which $\Gamma>3H$, also in this range of $C_\phi$
    the necessary condition for warm inflation $T>H$ is satisfied. For the case $m=1$
    ($\Gamma\propto T$),
    we have  obtained that the tensor-to-scalar ratio $r\sim 0$, and also
    we have found a lower limit for the parameter $C_\phi$ from the condition  $\Gamma>3H$.
 Finally,
 we have found that for the cases $m=0$ and $m=-1$, these warm-intermediate inflationary models
 are disproved from observational data, since the scalar spectral index $n_s>1$,
 then the models
  $\Gamma\propto \phi$ and $\Gamma\propto\phi^2/T$ do not
 work.

\begin{acknowledgments}
R.H. was supported by Comisi\'on Nacional de Ciencias y Tecnolog\'ia of Chile through
FONDECYT Grant N$^{0}$ 1130628 and DI-PUCV N$^{0}$ 123.724. N.V. was supported by Comisi\'on Nacional de Ciencias y Tecnolog\'ia of Chile through
FONDECYT Grant N$^{0}$ 3150490.
\end{acknowledgments}


\end{document}